\documentclass[twocolumn,showpacs,amsmath]{revtex4}
\usepackage{graphicx}%
\usepackage{dcolumn}
\usepackage{amsmath}
\usepackage{here}
\begin{document}

\title{Second-harmonic generation in vortex-induced waveguides}

\author{Jos\'e R. Salgueiro, Andreas H. Carlsson$^{\dagger}$, Elena Ostrovskaya, and Yuri Kivshar}

\affiliation{Nonlinear Physics Group, Research School of Physical
Sciences and Engineering, Australian National University, Canberra
ACT 0200, Australia}

\begin{abstract}
We study the second-harmonic generation and localization of light
in a reconfigurable waveguide induced by an optical vortex soliton
in a defocusing Kerr medium. We show that the vortex-induced
waveguide greatly improves conversion efficiency from the
fundamental to the second harmonic field.
\end{abstract}

\maketitle

Spatial optical solitons have become a topic of active research
promising many realistic applications and opening new directions
in nonlinear physics~\cite{book}. In its simplest form, a spatial
soliton is a single self-guided beam of a specific polarization
and frequency. Two (or more) mutually trapped components with
different polarizations or frequencies can form{\em a vector
soliton}.

One important application of spatial optical solitons is to induce
stable nondiffractive steerable waveguides that can guide and
direct another beam, thus creating a reconfigurable all-optical
circuit. The soliton-induced optical waveguides have been studied
theoretically and demonstrated experimentally in many
settings~\cite{wave1,wave2,wave3,wave4}. It was also shown that
soliton waveguides can be used for a number of important
applications, including the second-harmonic
generation~\cite{OL_lan}, directional couplers and beam
splitters~\cite{OL_guo}, and optical parametric
oscillators~\cite{OL_lan2}.

Dark solitons - localized dips on a background intensity - are more
attractive for soliton waveguiding applications because of their
greater stability and steerability~\cite{book}. Their
two-dimensional generalization, {\em optical vortex
solitons}~\cite{swartz}, may have a number of potential
advantages. Optical vortex solitons are light beams
self-trapped in two spatial dimensions and carrying a phase
dislocation. A systematic
analysis of the waveguides created by vortex solitons in a Kerr
medium~\cite{vort2,vort} demonstrates that an optical vortex can
guide both weak and strong probe beams, and that in the latter case
the vortex creates a stable vector soliton with its
guided component~\cite{book}.

In this Letter, we study the second-harmonic generation in a
reconfigurable vortex-induced waveguide and determine conditions
for significant enhancement of the conversion efficiency. We also
describe novel types of {\em three-component vector solitons}
created by a vortex beam together with both fundamental and
second-harmonic parametrically coupled localized modes guided by
the vortex-induced waveguide.

We consider two incoherently coupled beams with frequencies $\omega_0$ 
and $\omega_1$ propagating in a bulk nonlinear Kerr
medium. The $\omega_0$-beam propagates in a self-defocusing regime
and carries a phase dislocation. We assume that the phase-matching
conditions of the second-harmonic generation (SHG) are fulfilled
for the fundamental wave of frequency $\omega_1$ guided by the
vortex waveguide, so that it generates a second-harmonic (SH) wave
with the frequency $2\omega_1$. The SH wave is parametrically
coupled to the fundamental one and is also guided by the vortex
waveguide. Evolution of the slowly varying beam envelopes of the
vortex beam, the fundamental guided wave, and the SH wave can be
described by the following system of three coupled dimensionless
equations
\begin{equation}
\label{eq1}
   \begin{array}{l} {\displaystyle
      i \frac{\partial u}{\partial z} +\Delta_{\perp} u - (|u|^2 +
      \sigma |w|^2 + \rho |v|^2) u
      = 0,
   } \\*[9pt] {\displaystyle
       i \frac{\partial w}{\partial z} +\Delta_{\perp} w + w^*v -
      \sigma |u|^2 w
      = 0,
      } \\*[9pt] {\displaystyle
      2i \frac{\partial v}{\partial z} +\Delta_{\perp} v -\beta v + \frac{1}{2}w^2  -\rho
      |u|^2v
      = 0.
   } \end{array}
\end{equation}
where $u$, $w$, and $v$ are the normalized slowly varying complex
envelopes of the vortex beam, the fundamental field, and the SH
field, respectively. Other notations are: the Laplacian
$\Delta_{\perp}$ refers to the transverse coordinate ${\bf r}
=(x,y)$ measured in units of $r_0$, where $r_0^2 = 3\chi^{(3)}/16
\omega_1^2 [\chi^{(2)}]^2$ (see details in Ref.~\cite{ole}), $z$
is the beam propagation coordinate measured in units of
$z_0=2k_1r_0^2$. The parameter $\beta = 2z_0 \Delta k$ is
proportional to the wavevector mismatch $\Delta k = 2k_1 -k_2$,
whereas the nonlinear coupling coefficients $\sigma$ and $\rho$
are proportional to the corresponding third-order tensor
components~\cite{book}, and the self-action effects for the
fundamental and SH fields are neglected. Equations (\ref{eq1}) are
valid when spatial walk-off is negligible and the fundamental
frequency $\omega_1$ and its second harmonic are far from
resonance.

We emphasize that the model (\ref{eq1}) is the simplest of its
kind, which is most suitable for our feasibility study of SHG in
vortex-induced waveguides. It is clear that modelling of
particular experimental setups for realization of this concept
would require modifications of Eqs. (\ref{eq1}), according to the
geometry of an experiment and properties of nonlinear materials.
For example, in photorefractive crystals~\cite{OL_lan} one should take 
into account the nonlinearity saturation effect.

First, we analyze the stationary solutions of the model
(\ref{eq1}) in the form of the (2+1)-dimensional radially
symmetric nonlinear modes. We look for spatially localized
solutions in the polar coordinates $(r,\phi)$ of the form $u= u(r)
e^{-iz}e^{i\phi}$, $w=w(r)e^{i\lambda z}$, and $v=v(r)e^{i2\lambda
z}$, with the following asymptotic: $u(r) \rightarrow 1$, and
$(v(r), w(r)) \rightarrow 0$ for $r =\sqrt{x^2+y^2} \rightarrow
\infty$. Then, the mode amplitudes satisfy the system of 
$z$-independent equations
\begin{equation}
\label{eq2}
   \begin{array}{l} {\displaystyle
      \Delta_{r} u - \frac{1}{r^2} u + u - (u^2 + 2w^2 + 8v^2)u
      = 0,
   } \\*[9pt] {\displaystyle
      \Delta_{r} w - \lambda  w + wv - 2u^2 w
      = 0,
      } \\*[9pt] {\displaystyle
 \Delta_{r} v - (4\lambda + \beta)v +\frac{1}{2} w^2 - 8u^2 v
      = 0.
   } \end{array}
\end{equation}
where $\Delta_r = (1/r)d/dr(r d/dr)$ is the radial part of the
Laplacian, and for definiteness we have specified the parameters
of the cross-phase modulation interaction, $\sigma =2$ and $\rho
=8$. In Eqs. (\ref{eq2}), the real propagation constant $\lambda$
must be above cutoff, $\lambda > \lambda_c = {\rm max}(0,
-\beta/4)$, for $w$ and $v$ to be exponentially localized.

\begin{figure}[t]
\centerline{\includegraphics[width=3.0in]{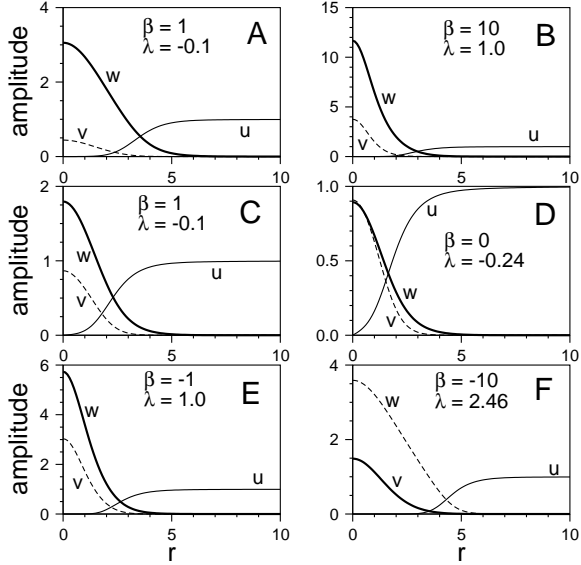}}
\caption{Spatial profiles of the three-wave vector soliton
components for the points A to F marked in Fig.~\ref{fig1}. Shown
are: the vortex amplitude $u(r)$ (thin solid), the fundamental
field $w(r)$ (thick solid), and the SH field $v(r)$ (dashed) at
the indicated values of $\beta$ and $\lambda$. } \label{fig2}
\end{figure}

\begin{figure}[t]
\centerline{\includegraphics[width=2.9in]{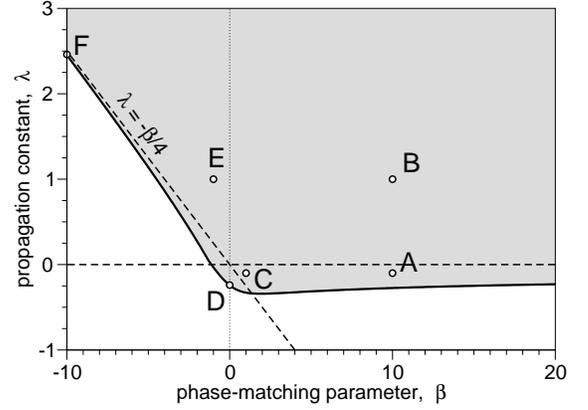}}
\caption{Region of existence (shaded) of the three-component
vector solitons of the model (\ref{eq1}) in the plane $(\lambda,
\beta)$. Marked points correspond to the localized modes shown in
Fig.~\ref{fig2}. } \label{fig1}
\end{figure}

Using the standard relaxation numerical technique, we find the
families of localized solutions of the system (\ref{eq2}) for
allowed values of $\beta$ and $\lambda$.  In Fig.~\ref{fig2} we
show several examples of the profiles of the three-wave localized
solutions for selected values of the parameters $\beta$ and
$\lambda$. The numerical results are summarized in Fig.~\ref{fig1}
which shows the existence domain as a shaded region of the plane
$(\lambda, \beta)$ with the boundary $\lambda = \lambda_{\rm th}$
(solid curve) found numerically and the asymptotic lines $\lambda
=0$ and $\lambda = -\beta/4$ (dashed) found from a simple analysis
of Eqs.~(\ref{eq2}).

All three-wave solutions of Eqs. (\ref{eq2}) can formally be
divided into {\em two categories} according to the dominant regime
of their formation: (i) vortex-waveguiding regime, $\lambda <0$,
and (ii) quadratic solitons regime, $\lambda >0$ . For
$\lambda_{\rm th} < \lambda < 0$, the parametrically coupled modes
$w$ and $v$ are localized only in the presence of the vortex, and
can be regarded as two guided modes of the effective
vortex-induced waveguide. Examples of such solutions are presented
in Fig.~\ref{fig2} by the cases A, C, and D. For $\lambda >0$, the
positive values of the propagation constant correspond to an
effectively self-localizing parametric nonlinearity acting between
the fields $w$ and $v$. These components can then become localized
even without the vortex, in the form of a parametric quadratic
soliton (see Fig.~\ref{fig2}, the cases B, E, and F). Parametric
coupling between the two fields is defined, as expected,  by the
value of the phase-matching parameter $\beta$.

In order to study the SHG process in the vortex-induced waveguide,
we employ the stationary solutions obtained above and analyze
numerically the evolution of the beams in the case when the SH
component is absent at the input. We perform all our calculations
for the case of {\em a finite-extent} input vortex beam, obtained
by superimposing the stationary vortex profile $u(r)$ onto a broad
super-Gaussian beam, $u_{\rm sG}= u(r) \exp [-(r^6/d]$, where $d=
10^8$. This form of initial conditions makes our predictions more
suitable for experimental verifications.

\begin{figure}[t]
\centerline{\includegraphics[width=3.4in]{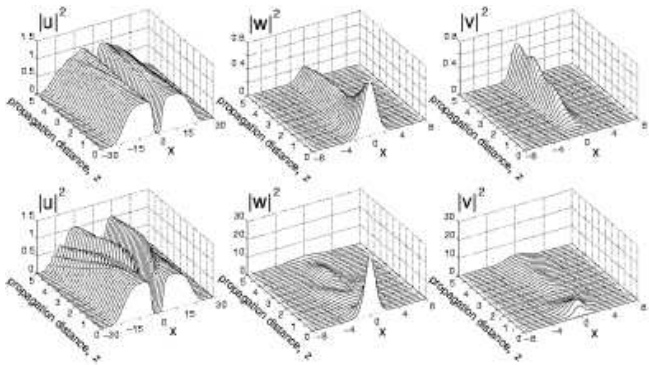}}
\caption{Examples of SHG in the vortex-induced waveguide with no
SH field at the input and the parameters corresponding to the
point D (upper row) and point E (lower row) in Fig.~\ref{fig2}
and \ref{fig1}. Notice the scale differences between the top and bottom rows.} \label{fig3}
\end{figure}

\begin{figure}[t]
\centerline{\includegraphics[width=3.0in]{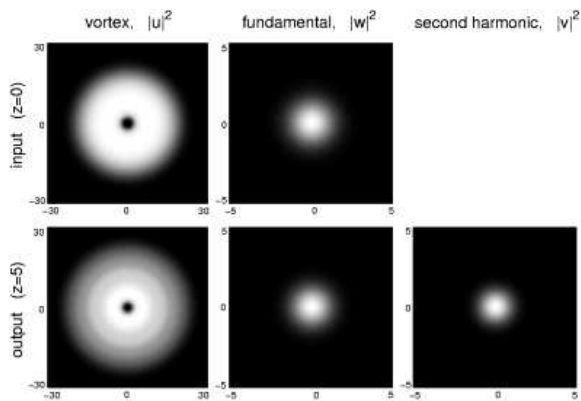}}
\caption{Grey-scaled images of the vortex waveguide and the guided
modes for the SHG process. Initial conditions correspond to a
vortex carried by a Gaussian beam and the fundamental wave, both
corresponding to the point D in Fig.~\ref{fig1}.} \label{fig4}
\end{figure}

The numerical results indicate that the generation of the SH field
from such an input differs dramatically for the vortex-waveguiding
and quadratic soliton regimes. Indeed, for $\lambda <0$ we observe
a good correspondence with the SHG theory. For large $\beta$, the
generated SH field is weak and the process corresponds to the
so-called nondepleted pump approximation in the SHG theory. Almost
perfect SHG is observed for $\beta$ close to zero, and in all such
cases the distortion of the vortex waveguide is weak.
Figure~\ref{fig3}(upper row) and Fig.~\ref{fig4} show an example
of the SHG process with $u(r)$ corresponding to the point D in
Fig.~\ref{fig1}. A good confinement of both fundamental and SH
guided modes can be seen with a very good conversion efficiency
and weak distortion of the vortex beam.

However, in the quadratic soliton regime, when $\lambda >0$, the
strong parametric interaction between the guided components does
not allow good energy conversion between the harmonics. Instead,
even for a high-intensity fundamental input, both the fundamental
and SH fields approach a stationary state with nonzero but
low-amplitude components. The SHG process becomes even worse for
the negative phase-matching. Figure~\ref{fig3}(lower row) shows an
example of a very strong mode coupling and vortex distortion
corresponding to the parameter region $\beta <0$ and $\lambda >0$.

If the vortex is removed at the input in the vortex-waveguiding
regime ($\lambda <0$), the SHG conversion efficiency drops by
at least one order of magnitude or more, and both the strong
fundamental and weak SH fields diffract rapidly. In the quadratic
soliton regime ($\lambda >0$), the effective self-focusing
nonlinearity of the second-order parametric interaction between
the fields $w$ and $v$ allows the formation of two-wave parametric
solitons even without the vortex component. However, in this case
the input power does not transfer into the SH field, it undergoes
a redistribution between the harmonics in such a way that both
fields either approach a stationary state corresponding to a
(2+1)-dimensional quadratic soliton (above the existence
threshold), or just diffract (below the threshold). To summarize,
our study of SHG in vortex-induced waveguides in different regimes
suggests that the enhanced conversion efficiency can be achieved
{\em only in the vortex-waveguiding regime}.

Possible experimental realizations of the concept of the SHG in
vortex-induced waveguides can be achieved in a crystal of
Fe:LiNbO$_3$ where phase-matching can be satisfied through the
birefringence effect at the angle $\theta = 81^{o}$ with respect
to the $z$-axis, provided the four-wave mixing effect is
suppressed. The other possibility is to employ photorefractive
crystals and the temperature tuning technique, similar to that reported
earlier~\cite{OL_lan}.

In conclusion, we have analyzed the simultaneous guidance of both
the fundamental and second-harmonic waves by an optical vortex
soliton. We have described novel classes of three-wave parametric
solitons with a vortex-soliton component, and have studied the
second-harmonic generation in the vortex-induced waveguides. For
the first time to our knowledge,  we demonstrated that larger
conversion efficiency of the SHG process can be
achieved in the vortex-waveguiding regime.

One of the authors (YK) thanks Ming-feng Shih and Solomon Saltiel
for useful discussions. The work was partially supported by the
Australian Research Council and the Secretar\'{\i}a de Estado de
Educaci\'on y Universidades of Spain through the European Social
Fund.

$^{\dagger}$ Currently at Acreo AB, 16440 Kista, Sweden.

\end{document}